\documentclass[%
 reprint,
 amsmath,amssymb,
 aps,
 onecolumn
]{revtex4-2}

\usepackage{graphicx}

\begin{document}

\title{Equilibrium Propagation for Learning in Lagrangian Dynamical Systems}

\author{Serge Massar}
\email{serge.massar@ulb.be}
\affiliation{Laboratoire d’Information Quantique CP224, Universit\'e libre de Bruxelles (ULB), Av. F. D. Roosevelt 50, 1050 Bruxelles, Belgium}

\begin{abstract}
We propose a  method for training dynamical systems governed by Lagrangian mechanics using Equilibrium Propagation. Our approach extends Equilibrium Propagation  -- initially developed for energy-based models --  to dynamical trajectories by leveraging the principle of action extremization. Training is achieved by gently nudging  trajectories toward desired targets and measuring how the variables conjugate to the parameters to be trained respond. This method is particularly suited to systems with periodic boundary conditions or fixed initial and final states, enabling efficient parameter updates without requiring explicit backpropagation through time. In the case of periodic boundary conditions, this approach yields  the semiclassical limit of Quantum Equilibrium Propagation. Applications to systems with dissipation are also discussed.
\end{abstract}

\date{March 2025}

\maketitle

\section{Introduction}

Most neural networks used  in machine learning, such as feedforward networks, are static rather than dynamic. By this we mean that  they
do not evolve dynamically during the inference phase. There are some exceptions such as recurrent neural networks (see e.g. \cite{mienye2024recurrent} for a recent review) and reservoir computing (see e.g. \cite{nakajima2021reservoir}), as well as neural differential equations  (see \cite{oh2025comprehensive} for a recent review) which themselves are connected to control theory  \cite{E2017,doi:10.1137/21M1411433} 

The most straightforward method to train the internal parameters of such time dependent neural networks  is backpropagation through time, see for instance \cite{hermans2015trainable} for its use in the case of reservoir computing.
However one may wish for more efficient training mechanisms which allow   all parameters to be updated at once. One may also wish for training methods adapted to the specific dynamical systems arising in classical mechanics, as these are particularly relevant for potential implementation in physical systems.
An interesting advance in this respect is \cite{PhysRevX.13.031020} in which dynamical systems governed by Hamiltonian dynamics are trained using both a forward and backward pass through the system, exploiting the time reversal invariance of the dynamics.

Here we show how dynamical systems governed by Lagrangian dynamics can be trained using the machine learning algorithm known as Equilibrium Propagation  \cite{SB17}, following the earlier work of  \cite{Scellier2021Thesis}.
Compared to \cite{PhysRevX.13.031020}, we note that Lagrangian mechanics is slightly more general, allowing for instance to incorporate some forms of dissipation. The main differences with respect to \cite{PhysRevX.13.031020} are  on the one hand that we do not need to explicitly propagate  the system backwards in time, but on the other hand our approach only  applies to certain classes of trajectories, such as periodic trajectories or trajectories with fixed initial and final conditions. 
After this article was submitted, a related work  was reported \cite{pourcel2025lagrangian} and a method for apply equilibrium propagation for damped linear systems with periodic trajectories was introduced in \cite{berneman2025equilibrium}.

Equilibrium Propagation, as formulated in \cite{SB17}, uses a neural network that minimises an energy functional. Reaching the minimum energy state takes time (either real time in a physical system, or computational time). Once the equilibrium state is reached, the training of the network is highly efficient. It is  obtained by nudging the output in the desired direction and measuring how the variables conjugate to the trainable parameters shift in response to the nudge. 
While reaching the minimum energy state is time consuming in digital implementations, it could be highly efficient in physical implementations that relax fast to their low energy configuration.
Conceptually, one expects that the Equilibrium Propagation approach can be used for training any machine learning system that is operated in a state that extremises a functional (not necessarily the energy). This was for instance the basis for extending Equilibrium Propagation to the 
quantum case\cite{massar2025equilibrium,scellier2024quantum,wanjura2024quantum} or to the thermal case \cite{SB17,massar2025equilibrium}. It is also the basis for the present generalisation to dynamical systems, the ``equilibrium" condition being here replaced by the condition that the equations of motion extremise the classical action.
Equilibrium Propagation was improved in \cite{L21} where symmetric nudging of the output that removes  biases and  improves performance was introduced. It was extended to vector field dynamics
\cite{scellier2018generalization}, to continuous time updates   \cite{ernoult2020equilibrium}, to spiking neural networks \cite{martin2021eqspike}, to large nudgings of the output \cite{laborieux2022holomorphic}. For a comparison with other energy based models, see \cite{Scellier2023EnergyBased}.

\begin{figure}
    \centering
        \includegraphics[width=0.9\linewidth]{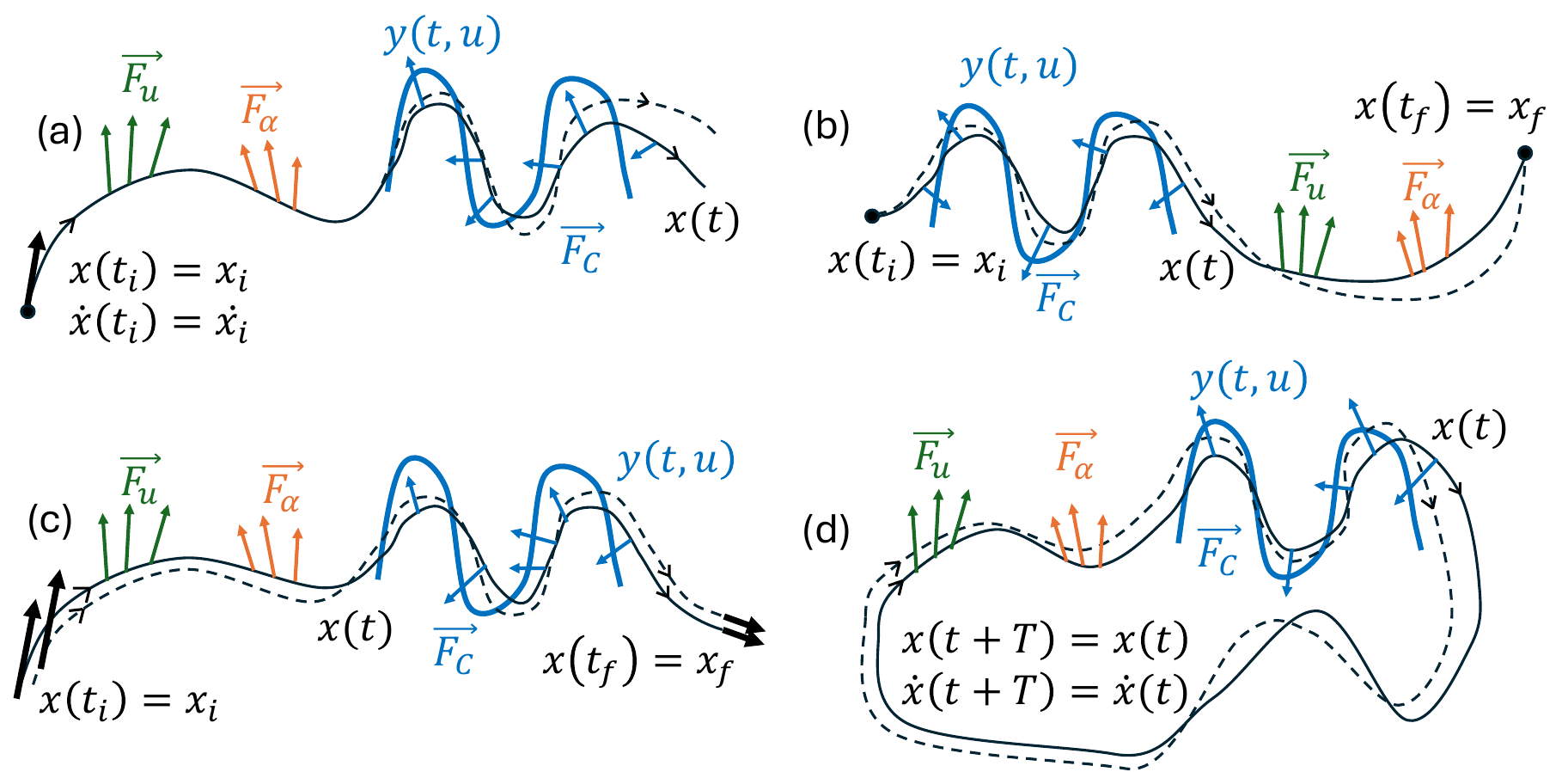}
    \caption{
Training a Lagrangian dynamical system: trajectories and forces. The black curve $x(t)$ is the trajectory. The  forces  $\vec F_u$ in green encode the input. The forces $ \vec F_\alpha$ in orange are used to train the system. 
The thick continuous blue line $y(t,u)$ is the target trajectory. The forces  $\vec F_C$ in blue are used to nudge the trajectory towards the target trajectory. These forces are defined in
section \ref{Sec:Example}.
The dashed black curve is the  trajectory after nudging towards the target $y(t,u)$, i.e. after $\vec F_C$ is applied.
In the figure, the forces apply in localised regions, for instance due to the time dependent functions  $\mu(t),\nu(t),\delta(t), \gamma(t)$ in the example of section \ref{Sec:Example}, or due to the $x$ dependence of the forces. Fixed initial and final positions are denoted by a thick black dot. Fixed initial or final velocities are denoted by thick black arrows.
 Panel (a) depicts the case in which one specifies the initial position $x(t_i)$ and velocity $\dot x (t_i)$. Equilibrium propagation does not apply in this case. Panel (b) (c) and (d) depict boundary conditions for wich Equilibrium Propagation can be applied. They are respectively fixed initial $x(t_i)$ and final $x(t_f)$ positions in Panel (b), fixed initial $\dot x(t_i)$ and final $\dot x(t_f)$ velocities in Panel (c), and periodic boundary conditions in Panel (d). 
  Note that in the figure we have denoted the forces using vector notation $\vec F$ and with multiple arrows to emphasize that they  act at mutliple locations along the trajectory. In the rest of the text, for compactness of notation, we drop the vector notation.
      }
    \label{fig:mainfigure}
\end{figure}

In Figure \ref{fig:mainfigure} we illustrate different examples.
An input is applied to the system, in the form of a force  $F_u(x,t)$.
The goal  is for the trajectory $x(t)$ to follow as closely as possible a target trajectory  $y(t,u)$.
This should be achieved by modifying parameters $\alpha$, and the corresponding 
 forces $ F_\alpha(x,t)$. The method of Equilibrium Propagation, adapted to the present context, consists of nudging the trajectory towards the desired trajectory using a force  $F_C(x,t)$, and using the modified trajectory to compute the update to $\alpha$. In the figure the nudged trajectory is depicted using a dashed line.
 
Consider first panel (a) which depicts a trajectory specified by its initial conditions, i.e.  
 initial position $x(t_i)$ and velocity $\dot x (t_i)$. In this case the method of Equilibrium Propagation does not apply.  This is clear from the illustrated example in which 
 the trajectory successively undergoes the  forces  $ F_u$ encoding input, 
the forces $ F_\alpha$ which  can be trained, finally undergoes the forces encoding 
which the desired trajectory  $y(t,u)$.  These forces act at successive times. It is obvious from the figure that nudging the trajectory towards the desired trajectory $y(t,u)$ cannot be used for training since the nudged trajectory is identical to the non-nudged trajectory in the region where the trainable parameters are applied.
 
Consider now panels (b,c,d) which depict boundary conditions for which Equilibrium Propagation can be applied. These boundary conditions are such  that applying a force someplace along the trajectory modifies the trajectory everywhere. This is necessary for Equilibrium Propagation, since  it is requires that
 nudging the trajectory towards the desired trajectory $y(t,u)$ also modifies the trajectory in the region where the trainable parameters are applied.
Panels (b,c,d) respectively depict fixed initial and final positions, fixed initial and final velocities, and periodic boundary conditions. 
  Note that the temporal order in which  the different forces, $ F_u$, $ F_\alpha$, $F_C$, are applied is not essential for Equilibrium Propagation. In fact these forces can  act simultaneously.

In Equilibrium propagation applied to dynamical systems, there is an intuitive way of understanding how to update the trainable force. Namely, if after nudging  the trajectory  so as to decrease the cost one finds that the trajectory has been displaced in the direction in which the trainable force $F_\alpha$ acts, then one should 
 update $\alpha$ so as to increase the trainable force. In other words one should update parameters so as to push the trajectory towards a better trajectory, i.e. towards the nudged trajectory. For instance, in panels (b) and (c)  the  nudged trajectory is displaced in the direction opposite to the trainable force $F_\alpha$ and one should therefore update $\alpha$ so as to decrease the trainable force, while in panel (d) one should increase the trainable force.
 This interpretation is derived and discussed at the end of Section \ref{Sec:3}.

 The most interesting boundary conditions for Equilibrium Propagation are probably the periodic boundary conditions, as these can most naturally arise in applications. Below we show  how they give rise to the semiclassical limit of Quantum Equilibrium Propagation. We also show how to incorporate dissipation in the formalism.
 
 In the remainder of this paper we first set up the problem and define our notations. Then we show how to apply equilibrium propagation to dynamical systems obeying Euler-Lagrange equations.

\section{Setting up the problem}\label{Sec:2}

\subsection{General formulation}

Let ${L}_0 (x,\dot x, t, \alpha, u ) $ be some Lagrangian depending on position
$x = (x_1, x_2, \dots, x_d) \in \mathbb{R}^d$, velocity $\dot x = (\dot x_1, \dot x_2, \dots, \dot x_d) \in \mathbb{R}^d$, time $t$, some parameters $\alpha$, and on the input $u$. 

Let $C(x,\dot x, t,  u ) $ be a cost density that measures how much the trajectory differs from the desired trajectory.
The total cost $ {\cal C}$ is the time integral of the cost density:
\begin{equation}
 {\cal C}(x,\dot x, t,u ) =  \int \! C(x, \dot x, t,u) \, dt
\end{equation}

The total Lagrangian is
\begin{equation}
{ L}  (x,\dot x, t, \alpha,\beta,u)  =  { L}_{0} (x,\dot x, t, \alpha, u )  +  \beta { C}(x,\dot x, t,  u )
\label{Eq:Lagrangian}
\end{equation}
where we have multiplied the cost by a  parameter $\beta$ which will be  used to  nudge the trajectory towards (or away from) the desired trajectory.
Although it is not essential for what follows, physical systems have a Lagrangian $L$ quadratic in $\dot x$.  The quadratic terms in $\dot x$ correspond to the kinetic energy,  linear terms in $\dot x$ can be interpreted as resulting from a magnetic field, and the term independent of $\dot x$ is minus the potential energy.

The Euler Lagrange equations are
\begin{equation}
\frac{d}{dt} \frac{\partial  {L} }{\partial {\dot x_j} } 
- \frac{\partial  {L} }{\partial {x_j} } = 0 \ .
\label{Eq:EL-equations}
\end{equation}
Their solution is denoted with  the subscript $*$ as
\begin{equation}
x_*(t,\alpha, \beta,u)\ .
\end{equation}
We also denote by the subscript $*$ quantities evaluated on the solution, for instance
 the Lagrangian evaluated on a solution is
\begin{equation}
{ L}_* (t, \alpha, \beta,u)= { L}  (x_*(t,\alpha,\beta,u),\dot x_*(t,\alpha,\beta,u), t, \alpha,\beta,u) \ ,
\end{equation}
while the cost on a solution is 
\begin{equation}
{\cal C}_*(\alpha, \beta,u)=\int \! C(x_*(t,\alpha,\beta,u),\dot x_*(t,\alpha,\beta,u),t,u) \, dt\ ,
\end{equation}
and the action evaluated on a solution (the on-shell action) is
\begin{equation}
{\cal  S}_* (\alpha, \beta,u) = \int  \!  { L}_* (t, \alpha, \beta,u) \, dt\ .
\end{equation}

The aim is, given a set of training examples, to adjust the parameters $\alpha$ in order to minimize the average cost (over the inputs $u$) when $\beta=0$:
\begin{equation}
\min_\alpha
\mathbb{E}_u\left[  {\cal C}_*(\alpha, \beta=0,u) \right]\ .
\label{Ex:ExpectC}
\end{equation}

\subsection{Illustrative example}\label{Sec:Example}

To be concrete, we consider an example. We suppose that ${L}_0= L_{S0}+D$ is the sum of two terms. The first term $L_{S0}$ describes a dynamical system unperturbed by the input:
\begin{eqnarray}
{ L}_{S0} (x,\dot x, t, \alpha) &=& \frac{m}{2} \sum_j \dot{x_j}^2 - k_0 \left( \frac{1}{2}\sum_j x_j^2 + \mu(t) \frac{1}{2}\sum_{j\neq k} \alpha_{jk}^{(2)} \rho(x_j)\rho(x_k)
+\nu(t) \sum_j \alpha_j^{(1)} \rho(x_j)
\right)\nonumber\\
&=& \frac{m}{2} \dot x^2 - k_0\left(\frac{1}{2} x^2  + \mu(t) \frac{1}{2}
\rho(x) \alpha^{(2)} \rho(x) 
+\nu(t) \alpha^{(1)} \rho (x)
\right)
\label{Eq:L0}
\end{eqnarray}
where 
$m$ is a mass, $k_0$ a spring constant,
$\mu(t)$ and $\nu(t)$ are time dependent functions,
$\rho$ is a sigmoid function such as 
$\rho(\cdot)=\tanh(\cdot)$. 
In this example there are $d(d+3)/2$ trainable parameters corresponding to the independent components of  the symmetric matrix $\alpha_{ij}^{(2)}$ and the $d$ components of the vector $\alpha_j^{(1)}$. We denote the set of trainable parameters by $\alpha= \{\alpha^{(2)}, \alpha^{(1)}  \}$. In the second line we have introduced the compact notation in which sums over indices are not written explicitly.

The last two terms in the lagrangian depend on the trainable parameters. The corresponding potential is noted $V_{train} (x,t,\alpha)=  \mu(t) \frac{1}{2}
\rho(x) \alpha^{(2)} \rho(x) 
+\nu(t) \alpha^{(1)} \rho (x)$. The corresponding force is denoted $F_\alpha = -\nabla V_{train}$, see Figure \ref{fig:mainfigure}.

The second term in the Lagrangian, which encodes  forces that depend on the inputs, is a $u$--dependent potential $D(t,x, u)$. The corresponding forces are denoted $F_u=-\nabla D $, see Figure \ref{fig:mainfigure}. 

For instance the input could depend on time dependent trajectories $z(t,u)$ generated by another dynamical system depending on the parameters $u$. The potential encoding the input could then take the form
\begin{eqnarray}
D(x, t,u) &=&\frac{k_{in}  \delta(t)  }{2} \sum_{j\in J_{\text{in}}  } ( x_j(t) - z_j(t,u) )^2\label{Eq:defD-2} 
\end{eqnarray}
where 
$J_{\text{in}} \subseteq \{1, \dots, d\}$ is
a subset of the dynamical variables, 
$\delta(t)$ is some time dependent function, for instance a gaussian or a constant, and $k_{in}$ a spring constant. The interpretation of the potential Eq. (\ref{Eq:defD-2})  is a spring  with time dependent spring constant $k _{in}\delta(t)$ connecting the two dynamical system $x$ and $z$.
The task could consist in classifying the trajectories $z(t,u)$ or in determining 
the parameters $u$ controlling the $z$ dynamical system.

As another example taken from more traditional machine learning tasks, suppose  the input consists of images drawn from the MNIST database. The grey value of each pixel  determines the force acting on the corresponding coordinate through the potential
\begin{eqnarray}
D(x, t,u) &=&  \frac{k_{in}  \delta(t)  }{2} \sum_{j \in J_{\text{in}}} ( x_j(t) - u_j )^2\label{Eq:defD} 
\end{eqnarray}
 The task in this case would be to classify the images.
Note that we can  clamp the inputs (by taking the spring constant $k_{in}$ to be arbitrarily large in Eq.  \eqref{Eq:defD}), so that the variables $x_j$, $j \in J_{\text{in}}$, are clamped  to the values $u_j$.

Our aim is for the trajectory to follow a target trajectory $y(t,u)$ that depends on the input $u$. To this end the
cost density $C(t,x, u) $, which measures how close the trajectory is to $y(t,u)$, is given by
\begin{equation}
 C(x, t,u) =  \frac{k_{out} \gamma(t) }{2} \sum_{j \in J_{\text{out}} }( x_j(t) - y_j(t,u))^2
\label{Eq:defC}
\end{equation}
where $J_{\text{out}} \subseteq \{1, \dots, d\}$ is the subset of the dynamical variables used for readout, $k_{out}$ is a spring constant,
and $\gamma(t)$ is some time dependent weight, for instance a gaussian. 
The corresponding force is denoted $ F_C =  -\nabla  C $, see Figure \ref{fig:mainfigure}.

For classification tasks $y(t,u)$ would naturally be taken to be time independent. For instance for the MNIST database, one would have $10$ outputs $\vert J_{\text{out}} \vert =10$, and one would use one-hot encoding. However the target output $ y_j(t,u)$ could also be a time dependent function, for instance  $y_j(t,u)= y_0 \sin (\omega(u) t)$ could be a sinusoidal function whose frequency $\omega(u)$ encodes the output.

Note that if the kinetic term is neglected (large mass limit), and the time dependent functions are all taken to be constant $\mu(t)=\nu(t)=\delta(t)=\gamma(t)=1$, then we recover identically the problem of \cite{SB17}. The novelty is that we now have a non trivial time dependence.

\section{Equilibrium Propagation}\label{Sec:3}

Let $\alpha_j$ be one of the parameters entering the Lagrangian. Given a training example (input $u$ and target $y(t,u)$) we wish to update the parameter $\alpha_j$ according to
\begin{equation}
\alpha_j \to \alpha_j - \lambda \frac{ d {\cal C}_*}{d\alpha_j }
\end{equation}
with $\lambda$  	a learning rate. Iterating this procedure will decrease the expected cost Eq. \eqref{Ex:ExpectC}.

Let us compute the derivative $\frac{d {\cal S}_* }{ d \beta}$ of the on-shell action with respect to $\beta$:
\begin{eqnarray}
\frac{ d {\cal S}_* (\alpha, \beta ,u)}{d \beta} 
&=& \int_{t_i}^{t_f} \! \left(
\frac{\partial { L}_*}{\partial \dot x}  \frac{ d \dot{x_*} }{d\beta }
+\frac{\partial { L}_*}{\partial  x}   \frac{ d x_* }{d \beta } 
+\frac{\partial { L}_*}{\partial \beta}\right) \, dt \nonumber\\
&=& 
\left[  \frac{\partial { L}_*}{\partial \dot x}   \frac{d x_* }{ d \beta }  \right]^{t_f}_{t_i} 
+\int_{t_i}^{t_f} \!   \left( \left(
-\frac{d}{d t}\frac{\partial {L}_*}{\partial \dot x}  
+\frac{\partial {L}_*}{\partial  x} \right) 
\frac{ d x_* }{ d\beta } 
+\frac{\partial { L}_*}{\partial \beta} \right)  \, dt 
\nonumber\\
&=& 
\left[  \frac{\partial {L}_*}{\partial \dot x}   \frac{d x_* }{d \beta }\right]^{t_f}_{t_i} 
+ {\cal C}_*
\label{Eq:dSdbeta}
\end{eqnarray}
where summation over the $d$ components of $x_*$ or $\dot x_*$ is understood as in Eq. \eqref{Eq:L0}.

Similarly, we have
\begin{eqnarray}
\frac{ d {\cal S}_* (\alpha, \beta ,u)}{d \alpha_j} 
&=& 
\left[  \frac{\partial {L}_*}{\partial \dot x}  \frac{d x_* }{d \alpha_j }  \right]^{t_f}_{t_i} 
+\int_{t_i}^{t_f} \! \frac{\partial {L}_*}{\partial \alpha_j}  \, dt \ .
\end{eqnarray}

We use the equality 
\begin{equation}
\frac{ d^2 {\cal S}_* (\alpha, \beta ,u)}{d \alpha_j d \beta}  = \frac{ d^2 {\cal S}_* (\alpha, \beta ,u)}{d \beta d \alpha_j }
\end{equation}
to obtain
\begin{eqnarray}
\frac{ d {\cal C}_* }{d \alpha_j} (\alpha, \beta ,u)
&=&
\frac{d}{d \beta} \int_{t_i}^{t_f} \! \frac{\partial { L}_*}{\partial \alpha_j}  \, dt 
+ \frac{d}{d \beta} \left[  \frac{\partial { L}_*}{\partial \dot x}   \frac{d x_* }{d \alpha_j}  \right]^{t_f}_{t_i} 
- \frac{d}{d \alpha_j} \left[  \frac{\partial { L}_*}{\partial \dot x}   \frac{d x_* }{d \beta }\right]^{t_f}_{t_i} \nonumber\\
&=&
\frac{d}{d \beta} \int_{t_i}^{t_f} \! \frac{\partial { L}_*}{\partial \alpha_j}  \, dt 
+  \left[  { B}_*
 \right]^{t_f}_{t_i} 
 \label{dCalpha-boundary}\\
{ B}_*(\alpha, \beta,u,t)
& = &
\frac{\partial^2 { L}_*}{\partial \dot x^2 }  \left(  
 \frac{d \dot x_* }{d \beta  } \frac{d x_* }{d \alpha_j }  
 - \frac{d \dot x_* }{d \alpha_j }   \frac{d x_* }{d \beta  } \right)
 + \left(  
\frac{\partial^2 {L}_*}{\partial \dot x \partial \beta  }   \frac{d x_* }{d \alpha_j } 
 - \frac{\partial^2 {L}_*}{\partial \dot x \partial \alpha_j  } 
  \frac{d x_* }{d \beta  } \right)
   \label{B-boundary}
\end{eqnarray}
where we have used the symmetry of the matrices 
$\frac{\partial^2 {L}_*}{\partial  x_j \partial \dot x_k  } $ and
$\frac{d x_{*_j} }{d \alpha_j }  \frac{d x_{*_k} }{d \alpha_j }  $ to simplify the expression of the boundary term Eq. \eqref{B-boundary}.

Let us suppose that the boundary term $ \left[  {B}_*
 \right]^{t_f}_{t_i} $ vanishes  (we discuss how to impose this condition in the next section).
Then we can estimate the derivative of the cost function by finite differences.
We compute two trajectories  $x_*(\pm \beta)$ which are nudged by the small positive and negative values of $\beta$--dependent cost term. From these trajectories, we can obtain the derivative 
$\frac{ d {\cal C}_* }{d \alpha_j}$ by evaluating through integration how the variable conjugate to $\alpha_j$, i.e. $ \int_{t_i}^{t_f} \! \frac{\partial {L}_*}{\partial \alpha_j}  \, dt $, changes when we change the cost. We thus obtain
\begin{equation}
\frac{ d {\cal C}_* }{d \alpha_j}
\approx
\frac{1}{ 2 \beta} \left( \int_{t_i}^{t_f} \! \frac{\partial { L}_* (+\beta)}{\partial \alpha_j}  \, dt 
-
\int_{t_i}^{t_f} \! \frac{\partial { L}_* (-\beta)}{\partial \alpha_j}  \, dt \right)\ .
\label{dCdalpha}
\end{equation}
This is the analog for dynamical systems of the Equilibrium Propagation update rule  \cite{SB17, L21}, as previously derived in \citep{Scellier2021Thesis}. Note the economy of the method: from only two trajectories, one nudged towards the desired trajectory ($\beta$ positive) and one nudged in the opposite direction ($\beta$ negative) we obtain all the gradients, and hence can update all the parameters.

To interpret Eq. (\ref{dCdalpha}), we consider the simple case where the only dependence of the Lagrangian on $\alpha_j$ is through a potential energy term linear in $\alpha_j$ as in the example Eq. (\ref{Eq:L0}). That is we suppose that the relevant potential energy term has the form $V_{train}(x,t,\alpha)=\sum_j \alpha_j v_{j}(x,t)$. 
The associated force is $F_\alpha = \sum_j \alpha_j  f_{\alpha_j}$ with $f_{\alpha_j}=-\nabla v_{j}$ the force exerted per unit $\alpha_j$. Then $\frac{\partial L}{\partial \alpha_j}=-v_j(x,t)$ and we have
\begin{equation}
\frac{ d {\cal C}_* }{d \alpha_j}
=
\int_{t_i}^{t_f} \!  -f_{\alpha_j *} \frac{\partial x_*}{\partial \beta} \, dt \ .
\label{dCdalpha-example}
\end{equation}

The interpretation  of Eq. (\ref{dCdalpha-example}) is very natural and intuitive. If we nudge the trajectory so as to reduce the cost, and we find that the change in the trajectory $\frac{\partial x_*}{\partial \beta} $ is in the direction of the trainable force $\alpha_j  f_{\alpha_j}$, then we should increase $\alpha_j$, i.e. increase the force so as to push the trajectory closer towards the desired trajectory. Conversely we should reduce $\alpha_j$ if reducing the cost has nudged the trajectory in the opposite direction to the trainable force $\alpha_j  f_{\alpha_j}$.

Finally, recall that work is the product of force times displacement. The term $\frac{\partial x_*}{\partial \beta} $ is the diplacement induced by nudging the trajectory with the cost term, per unit of nudging $\beta$. Hence the integrand in 
Eq. (\ref{dCdalpha-example}) is minus the work induced by nudging the trajectory with the cost term, per unit of nudging $\beta$ and per unit $\alpha_j$.

\section{ Vanishing of boundary terms}\label{Sec:4}

\subsection{Causal boundary conditions do not work}

Equation \eqref{dCdalpha} is only valid if the boundary terms in Eq. \eqref{dCalpha-boundary} vanish. We now discuss when this is the case. This requires imposing appropriate  boundary conditions.

Note that  the usual   boundary conditions, in which the initial position $x(t_i)$ and velocity $\dot x(t_i)$ are specified, do not work.
It can even be the case that the first term in Eq. \eqref{dCdalpha} vanishes and the gradient $d {\cal C}_* / {d \alpha_j}$ is entirely given by the boundary terms. This is, for instance, the case in the example given in Figure \ref{fig:mainfigure} panel (a) where the right hand side of Eq. \eqref{dCdalpha} vanishes since, due to the order in which $F_\alpha$ and the nudging towards $y(t,u)$ are applied,
$\frac{\partial { L}_* }{\partial \alpha_j}$ does not depend on $\beta$.

\subsection{ Fixed initial and final position}\label{SubSec-fixedPos}
The simplest boundary condition that ensures that the boundary terms in Eq. \eqref{dCalpha-boundary} vanish is to fix both the initial and final positions:
\begin{eqnarray}
x_*(t_i)&=&x_i \nonumber\\
x_*(t_f)&=&x_f \ .
\end{eqnarray}
This case is depicted in Figure  \ref{fig:mainfigure}  Panels (b). It was introduced in \cite{Scellier2021Thesis}.

\subsection{ Fixed initial and final velocity }\label{SubSec-fixedVel}

Another possible boundary condition  is to fix both the initial and final velocity. This ensures that the first boundary term in Eq. \eqref{dCalpha-boundary} vanishes (since then $d \dot x_* / d \beta = d \dot x_* / d \alpha_j = 0$ on the boundary). However an additional condition must be added so that the second boundary term vanishes, namely the 
dependence of Lagrangian on $\beta$ and $\alpha_j$ must vanish at the initial and final time as expressed in the following  conditions
\begin{eqnarray}
\dot x_*(t_i)&=&\dot x_i \nonumber\\
 \dot x_*(t_f)&=&\dot x_f \nonumber\\
\frac{\partial^2 { L}}{\partial \beta  \partial \dot x  }  (t_i) &=&
\frac{\partial^2 {L}}{\partial \beta \partial \dot x  }  (t_f) = 0\nonumber\\
\frac{\partial^2 { L}}{ \partial \alpha_j  \partial \dot x  } (t_i) 
&=&\frac{\partial^2 { L}}{ \partial \alpha_j  \partial \dot x  } (t_f)  = 0  \ .
\end{eqnarray}
The last two conditions are easy to implement: they hold  when the cost  and terms in the Lagrangian that depend on the trainable parameters $\alpha$ contain no dependence on the velocities, and they also hold when the cost  and terms in the Lagrangian that depend on $\alpha$ vanish at the intial and final time $t_i$ and $t_f$. This case is depicted in Figure  \ref{fig:mainfigure}  Panels (c).

\subsection{ Periodic boundary conditions}

\subsubsection{General condition}

Particularly interesting are periodic boundary conditions:
\begin{eqnarray}
x_*(t_i)&=&x_*(t_f)\nonumber\\
\dot x_*(t_i)&=&\dot x_*(t_f)\nonumber\\
 { L} (x,\dot x, t_i)&=&  { L} (x,\dot x, t_f)\ .
\end{eqnarray}

It is useful to rephrase them by extending the Lagrangian and trajectories to periodic functions, using  $t_f = t_i + T$, as
\begin{eqnarray}
 { L} (x,\dot x, t + T)&=&  { L} (x,\dot x, t)  \quad \text{for all $t$,}\label{Eq:PeriodT}\\
 x_*(t + T)&=&x_*(t)  \quad \text{for all $t$.} \label{Eq:PeriodT-x} 
\end{eqnarray}
This is illustrated  in Figure  \ref{fig:mainfigure}  Panels (d).

\subsubsection{Time independent case and Hamiltonian formalism}

Before analysing in more detail periodic boundary conditions, we recall the Hamiltonian formulation of the equations of motion.
The momentum is defined by
\begin{equation}
p_i(x,\dot x, t) = \frac{\partial { L} (x,\dot x, t)  }{\partial \dot x_i} \ .
\end{equation}
The energy is defined by
\begin{equation}
{ E}  (x,\dot x, t)  = \sum_i  p_i  (x,\dot x, t)  \dot x_i -  { L}  (x,\dot x, t) \ .
\end{equation}
The action can then be rewritten as
\begin{equation}
{\cal S} = \int \!  \sum_i  p_i \dot x_i -  { E} \, dt\ .
\end{equation}

Suppose that the Lagrangian does not have any explicit time dependence. Then
we can impose $ x_*(t + T)=x_*(t)$ for all $t$, but leave $T$ a free parameter, and 
  impose other constraints. 
We consider three such constraints.  

{\bf Constant solution.}
We can take the solutions to be constant $\dot x_* = 0$, in which case the system is at a local minimum of the energy. We then recover the original problem of \cite{SB17}, since the time dependence has dissapeared.

{\bf Fixed energy.}
We can take the solutions to be periodic with fixed energy $E$. In this case the period $T$ is not fixed, and will adjust itself in order to satisfy the condition on the energy.

{\bf Fixed reduced action.}
We can take the solutions to be periodic with fixed reduced action 
 \begin{equation}
 {\cal S}_0 = \oint \!  \sum_i  p_i \dot x_i \, dt =  \oint \!   p  \, dx
 \end{equation}
In this case the period $T$ is not fixed, and will adjust itself in order to satisfy the condition on $S_0$.

Imposing fixed reduced action $S_0$ can be viewed as the semiclassical version of   Quantum Equilibrium Propagation \cite{massar2025equilibrium,scellier2024quantum,wanjura2024quantum}. Indeed the semi-classical quantization conditions is
 \begin{equation}
\oint \!   p  \, dx = 2 \pi (n + \alpha) \hbar
 \end{equation}
 with $n$ integer and $\alpha$ the Maslov index of the trajectory \cite{Percival1977}.

\subsection{Damped equations of motion}

If the equations of motion include damping, then they are no longer symmetric with respect to time reversal. There are several methods to introduce damping in the Lagrangian formalism, see e.g. \cite{riewe1996nonconservative}. 
Here we consider a specific method, and consider how it affects equilibrium propagation.

To include damping, we  take the original  Lagrangian Eq. \eqref{Eq:Lagrangian} and multiply it by $e^{\Gamma t}$ to obtain
\begin{equation}
\tilde { L}  (x,\dot x, t, \alpha,\beta,u)  =e^{\Gamma t}  { L}  (x,\dot x, t, \alpha,\beta,u) \ .
\label{Eq:Lagrangian-Gamma}
\end{equation}
The new equations of motion are the old equations of motion plus  a damping term:
\begin{equation}
\frac{d}{dt} \frac{\partial  { L} }{\partial {\dot x_i} } + \Gamma \frac{\partial  {L} }{\partial {\dot x_i} }
- \frac{\partial  { L} }{\partial {x_i} } = 0 \ .
\label{Eq:EL-equations-Damped}
\end{equation}

If we apply fixed initial and final position, or fixed initial and final velocity, as in Subsections \ref{SubSec-fixedPos} and \ref{SubSec-fixedVel}, we can apply Equilibrium Propagation as before. The only difference is that the time dependent functions $\gamma(t)$ will appear in Eq. \eqref{dCdalpha-example}.

Let us now consider periodic boundary conditions. We suppose that the original Lagrangian $ { L} $ is periodic in time with period $T$, see Eq. \eqref{Eq:PeriodT}. However with damping the boundary terms  no longer vanish because the new Lagrangian 
$\tilde { L} $ is not periodic in time.
However, if the damping is weak, the boundary terms will be small, as we now show.
Suppose $x_*$ is a solution of the
 the damped equations of motion Eq. \eqref{Eq:EL-equations-Damped} with 
 periodic boundary conditions,  Eqs. \eqref{Eq:PeriodT} and \eqref{Eq:PeriodT-x}.
 
One finds that Eq. \eqref{dCalpha-boundary} becomes:
\begin{eqnarray}
\frac{ d \tilde {\cal C}_* }{d \alpha_j} 
&=&
\frac{d}{d \beta} \int_{0}^{T} \! \frac{\partial \tilde  { L}_*}{\partial \alpha_j}  \, dt 
+  (e^{\Gamma T}-1)  { B}(0)\\
&\approx & 
\frac{d}{d \beta} \int_{0}^{T} \! \frac{\partial \tilde  { L}_*}{\partial \alpha_j}  \, dt 
+  \Gamma T  { B}(0)\label{Eq:smallDamping}
\end{eqnarray}
where $ { B}(t)$ is the boundary term Eq. \eqref{B-boundary} computed for the original Lagrangian $ { L} $, and
where in Eq. \eqref{Eq:smallDamping}  we have supposed that the damping is small $\Gamma T \ll 1$. We see that as the damping decreases, the error made if one neglects the boundary terms becomes smaller.

\section{Conclusion}

In this work, we have generalized the framework of Equilibrium Propagation to the training of dynamical systems governed by Lagrangian mechanics, with the cost expressed as a functional that can be included in the Lagrangian.  By exploiting the structure of the underlying dynamics, we derived an efficient method to compute the gradient of the cost function with respect to the trainable parameters.
This approach  requires only two perturbed trajectories to estimate all parameter updates.

Importantly, we showed that successful application of this method hinges on the choice of appropriate boundary conditions such as periodic trajectories or trajectories with fixed initial and final states. In particular, periodic boundary conditions are especially attractive, as they naturally connect to the semiclassical limit of quantum Equilibrium Propagation and can accommodate weak dissipation effects. We note that an alternative method for applying equilibrium propagation to damped linear systems with periodic trajectories was introduced in \cite{berneman2025equilibrium}

Our results show that Equilibrium Propagation can be extended beyond 
static energy--based models to a broad class of dynamical systems, including those with time--dependent or dissipative effects. This opens new avenues for  training physical, analog, neuromorphic computing systems, where the natural dynamics of the system can be harnessed for learning.

Future work may explore the implementation of these ideas in  physical systems, the extension to stochastic dynamics, and connections with Hamiltonian-based approaches and control theory.

\acknowledgments

I  thank Bortolo Mognetti and Dimitri Vanden Abeele for very helpful discussions. I acknowledge financial support by the Fonds de la Recherche Scientifique -- FNRS, Belgium under   EOS project 40007536.

\bibliography{EPDDynSyst}

 \end{document}